# Coherent transport in extremely underdoped Nd$_{1.2}$Ba$_{1.8}$Cu$_3$O$_z$ nanostructures.

F Carillo[1,*], G M De Luca[2], D Montemurro[1], G Papari[1], M Salluzzo[2], D Stornaiuolo[2], F Tafuri[2,3], and F Beltram[1]
1. NEST-Scuola Normale Superiore and Istituto Nanoscienze – CNR, Piazza S. Silvestro 12, I-56127 Pisa.
2. CNR-SPIN Complesso Univ. Montesantangelo, Via Cinthia, 80125 Napoli
3. Dipartimento Ingegneria dell'Informazione, seconda Università di Napoli, Aversa.

* E-mail: franco.carillo@sns.it

**Abstract.** Proximity-effect and resistance magneto-fluctuations measurements in submicron Nd$_{1.2}$Ba$_{1.8}$Cu$_3$O$_z$ (NBCO) nano-loops are reported to investigate coherent charge transport in the non-superconducting state. We find an unexpected inhibition of cooper pair transport, and a destruction of the induced superconductivity, by lowering the temperature from 6 K to 250 mK. This effect is accompanied by a significant change in the conductance-voltage characteristics and in the zero bias conductance response to the magnetic field pointing to the activation of a strong pair breaking mechanism at lower temperature. The data are discussed in the framework of mesoscopic effects specific to superconducting nanostructures, proximity effect and high temperature superconductivity.

## 1. Introduction

Recent progresses in the application of nanotechnology to high critical-temperature superconductors (HTS) have opened the way to nanoscale-transport experiments within these systems [1-4]. In particular, coherence [3, 4], pairing-symmetry details [5, 6], and the possible coexistence in momentum as well as in real space of different quantum phases [7-9], could profoundly influence transport phenomena in HTS nanostructures. However, the existence of intrinsic inhomogeneous phase is still controversial [10, 11] and the study of HTS nano-devices, especially in the extreme underdoped region, would contribute with important information on the physics at this length scales. Additionally, nanoscale devices allow the study of coherent transport of quasiparticles in the normal states of cuprates, for which there are only few reports [12]. Recently, we demonstrated the possibility to study coherence in a single YBCO nano-loop in the superconducting state [3].

In this paper Nd$_{1.2}$Ba$_{1.8}$Cu$_3$O$_z$ nanodevices are used to study coherent transport in the normal state of severely-underdoped HTS films. The experiment was designed to take advantage of a nanoloop interferometer geometry and was based on the accurate control of the doping level in the underdoped phase. When transport is constrained in a nanostructure of a size comparable to the spatial charge-correlation length [13], transport properties are more sensitive to inhomogeneity in the electronic phase [1], since they are less affected by averaging effects. We show that a non-superconducting nanoloop (Fig 1a) with a doping level below 0.06± 0.01 holes/CuO$_2$ plane, therefore, at the frontier between superconducting-insulating region of the phase diagram, can sustain coherent transport of pairs between two superconducting electrodes by proximity effect. This induced superconductivity displays an unexpected reentrant behavior at low temperatures that signals the inhibition of Cooper-pair transport by a strong pair breaking effect, but preserving coherent transport of quasiparticles in the nanoloop.

## 2. Description of the experiment

In this study we employed 60 nm thick untwinned c-axis $Nd_{1.2}Ba_{1.8}Cu_3O_z$ thin films [14] characterized by extremely smooth surfaces with an rms roughness of 1.0 nm (i.e. 1 unit cell) on 1x1 $\mu m^2$. We chose untwinned films of high structural quality to preserve the orientation of d wave order parameter on the whole substrate [15]. Moreover, these films are characterized by a 2D-insland growth mode, which result in extremely smooth surfaces and in a reduced influence of the grain boundaries in the transport properties [16]. Optimally oxygenated Nd-rich NdBCO films are underdoped, with a critical temperature $T_C$ = 65 K and number of carriers of 0.12± 0.01 holes/unit cell, estimated from Seebeck effect [14]. We adjusted the oxygen content by appropriate annealing procedures in a reducing atmosphere. By adjusting the annealing time and by using different reducing atmosphere (typically different $Ar/O_2$ ratios) we were able to control the Tc of our films from the maximum of 65 K down to the non-superconducting state. Previous investigations showed that the $T_C$ and the number of carriers of our films are correlated [14], following the classical universal curve valid for 123 cuprate superconductors, as shown in Fig. 1b. Thus in this experiment we have studied samples with a doping above and below 0.06 holes/$CuO_2$ plane, allowing us to investigate the transport across the superconducting-insulating transition.

The main steps for the fabrication procedure of the HTS nanodevices can be found in Ref. 3. The critical temperature of oxygen-reduced samples is around 35 K (estimated doping of 0.08 holes/unit cell) for large wires and about 5 K for 250-nm-wide wires. In agreement with our previous results on YBCO [3], $T_C$ of submicron bridges decreases when reducing the line width and, in this sample, structures thinner than 200 nm remain non-superconducting even at the lowest temperature tested (250 mK).

Doping level and disorder in the ring are two distinct sources for the reduction of $T_C$ in nanowires [17,18]. It is difficult to evaluate separately these two factors. Based on the experimental value of $T_C$ we can estimate a *nominal* doping level for the nanostructure. This represents the doping level of an ideal structure without disorder and will be used as an effective parameter labeling the specific device and indicating its overall electronic properties. Of course these estimates can be done only for those devices, or parts of the circuit (e.g. leads), showing a $T_C$. For devices which remain non superconducting we can only estimate an upper-bound for the doping level (< 0.06 holes /unit cell).

All measurements presented in this work were performed in a $^3$He cryostat equipped with PID-controlled temperature and a superconducting magnet. The cryostat was shielded from external magnetic fields by nested cryoperm, lead, and Nb foils and was equipped with a double-stage of copper-oxide and RC filters thermally anchored at 1.5 K. Magnetoresistance measurements were performed in the four-probe configuration and by standard lock-in techniques using an AC excitation current of 30 nA at 17 Hz. Temperature fluctuations during the measurement were < 0.5% and could induce only marginal measurement distortions on the same scale.

## 3. Discussion

Fig. 1a reports a SEM micrograph of one of the two nano-loops discussed here (Ring A). The ring has an average radius of 220 nm, arm width of 170 nm and 380 nm wide leads. Fig. 2a reports the resistance vs. temperature curve and shows that at 40 K the transition of wider connecting leads (W >> 1 um) occurs. The overall temperature dependence of the resistance is plotted in the inset and shows a behavior characteristic of severely underdoped samples [14]. The superconducting transition of the device is not complete even at 250 mK (Fig. 2a ). This implies that the nanoring has a doping lower than 0.06 holes/unit cell and is not superconducting. Conversely, the electrodes undergo a complete transition at T larger than 4.2K since they are much wider (>200nm); thus, below 4.2 K the narrower region of the ring plays the role of a normal barrier. The overall superconductor-normal-superconductor (SNS) geometry of the device is confirmed by transport data. Above 2 K, the conductance increases at low voltages as expected for the coherent flow of Cooper pairs and quasi-particles in the nanoring (Fig. 2b) [19-21]. Cooper pairs injected from the superconducting electrodes can diffuse into the nanoring and lead to a zero bias peak (ZBP) due to a vanishingly small supercurrent through the ring [19-21]. Superimposed onto a quasi-parabolic background, relatively regular oscillations of the resistance as function of magnetic field reproducibly appeared both at 250mK and at 4.2K (Figure 3a), consistently with what is commonly associated with coherent transport through a ring [3].

The increase of the zero bias resistance of the device below 3 K (Fig. 2a) indicates that the nanoring is entering a different transport regime. At lower temperatures, somewhat unexpectedly, we observe a significant change in the conductance vs. voltage ($G(V)$) behavior: the excess current at low voltages appearing in the ZBP decreases when lowering the temperature and a minimum in the conductance appears at zero bias (zero bias dip, ZBD) [Fig. 2b]. In conventional SNS systems, a temperature decrease produces an enhancement of the conductance and may even lead to a measurable supercurrent for sufficiently short barrier distances [19-21]. Our data display the opposite behavior: a reduction of Cooper-pair transport at lower temperature. At large voltages (>1 mV) the conductance remains unchanged at all temperatures. This suggests that the increase of the zero bias resistance is not related to a change of the normal state properties of the ring, but is driven by Cooper-pair flow.

Reduction of superconductivity with decreasing temperatures was previously reported by other authors [22,23]. A reentrant behavior of R(T) was observed by Stewart et al. [22] on nanopatterned Bismuth film and was attributed to the localization of Cooper pairs triggered by disorder. We cannot exclude the possibility that the level of disorder is large enough to produce pair localization also in our nanostructures.

Inhibition of Cooper-pair transport at low temperatures was also reported for InAs-based two-dimensional electron gases (2DEG) coupled to niobium electrodes [22]. In that case, the effect was attributed to the localization of single electrons in the 2DEG. The latter phenomenon occurred for large ratios $R_\square/R_Q \sim 30$ between the normal square resistance ($R_\square = R_N L/W$) and the quantum of resistance ($R_Q = h/2e^2$). In our case the $R_\square/R_Q$ ratio for a single $CuO_2$ bilayer. is about $R_\square/R_Q = 3.4$ [24] This value is an upperbound of the real $R_\square/R_Q$ per $CuO_2$ plane. As matter of fact, the effective width and thickness of the device contributing to the transport are much smaller that the geometric one. This is mainly due to the effect of Argon ion milling (during the fabrication procedure) which might reduce the number of conducting $CuO_2$ planes and might produce a damaged region at the edge of the wire. This effects are evidenced by the apparently too large normal state resistivity found in our nanostructures (about 2.3 mΩ cm for ring A and 1.4 mΩ cm for ring B ), but similar to those found in other HTS nanostructures [1,3].

In order to further investigate the phenomenology reported above, additional measurements were performed on a larger loop (Ring B) with average radius of 350 nm and branch width of 260 nm, see Inset of Figure 4-d). Ring B and Ring A have been fabricated at the same time on the same chip. They have also been measured during the same measurement session with the same equipment. The only difference between the two devices is their size. For ring B we found the usual monotonic enhancement of conductance while decreasing temperatures (inset of Fig. 4, panel a). Magnetoresistance oscillations were analogous to those observed in Ref. 3 with no signatures of *h/e* components, as also occurring in arrays of nanoloops [4]. The large amplitude of magnetoresistance oscillations in Ring B at temperatures much lower than its $T_C$ can be attributed to the flux flow in the branches of the ring as recently demonstrated in arrays of HTS nanoloops [4]. The behavior of Ring B does suggest that the unconventional behavior of the smaller ring (Ring A) is indeed related to its size. Further insight was obtained by measuring the magnetoresistance of ring A. In order to identify the relative contributions of both Cooper pairs and quasiparticles to the current, we applied the Fast Fourier transformation to the ΔR(H) curves of Fig. 3a after the subtraction of the parabolic background. The results are shown in Figure 3b together with the expected ranges of the FFT components associated to h/2e (Cooper pairs, cyan bar) and to h/e (quasiparticles, green bar) periodicity: they were calculated using the maximum and minimum radius of the ring, as measured from the SEM image. The dominant components in the FFT are the same at both temperatures, and this implies that the periodicity of the magnetoresistance curves does not substantially change with the temperature. This frequency distribution of FFT components is very different from what previously observed in Ref. 3 for superconducting nanoloops and for Ring B in this work (Fig. 4d). The concentration of the FFT peaks at the center of h/2e bar (see Fig. 4d and ref 3) is consistent with a preferential flow of Cooper pairs in the central region of the two branches (of the ring) where disorder and loss of oxygen induced by the etching are presumably lower and superconductivity is stronger. In the smaller ring (Ring A), we observe just a minor peak at the center of h/2e bar (~84 1/T) and only at 4.2K. This seems to suggest coherent motion of Cooper pairs at 4.2K, fully consistent with the presence of excess conductance in G(V) at zero bias at T>2K. Similarly, at 250 mK the FFT components in the central part of h/2e are absent and there is a ZBD in the *G(V)* (Fig. 2b).

The detection of Cooper-pair flow from the periodicity of magnetoresistance oscillations was already established in experimental [25-26] and theoretical [27] studies on metallic (non-superconducting) nanorings connected to conventional superconducting leads. The zero-bias resistance modulates with a prevailing h/2e or h/e periodicity depending on the superconducting state of the leads. In those systems, magnetoresistance oscillations displayed an h/e periodicity when the leads were in the normal state, i.e. above $H_C$ or $T_C$, signaling coherent transport of electrons. Differently, in our system the switching off of Cooper pairs flow can not stem from the suppression of superconductivity in the leads, since they remain superconducting in the whole temperature range explored, but must be linked to a change in the transport properties of the nanoloop itself. The data indicate that at low temperatures the ring enters a state that inhibits Cooper pair flow but not single-electron coherent transport. Single electron transport was already suggested by Doiron-Leyraud et al [28] to explain thermal conductivity at low temperature in underdoped non-superconducting YBCO crystals.

Even if some caution is required when interpreting magnetoresistance data to evaluate the influence of possible contributions to the magnetoresistance from the leads [29], the internal consistency between magnetoresistance and trans-conductance data is a robust argument. From a microscopic point of view, an inhibition of Cooper-pair flow could be due to the switch on, at low temperature, of strong scattering centers capable to break cooper pairs and reduce the superconducting correlation in the ring. In a HTS picture a suggestion on the nature of this strong scattering centers comes from experiments on muon spin rotation, where the existence of a static magnetic order is demonstrated at low temperatures for underdoped cuprates [7,8]. In this framework the formation of nanodomains of different electronic phases (e.g. spin order) could act as an effective cooper pair breaker, analogously to the presence of magnetic impurities in conventional SNS systems [20]. Nevertheless our results are not necessarily to be interpreted in terms of the existence of nano-domains of different electronic phases. Other pair breaking mechanism could be good candidates. For example the magnetic order of Nd spins at low temperature (<2K) [30] could effectively act as pair-breaker preserving the coherence of single charge carrier.

In principle the h/e signal found on Ring A could be the effect of D-wave symmetry and of the long coherence of quasi particle in the nodes [6]. In fact Loder et all. [5] predicted an *h/e* component in the periodicity of mesoscopic loops of superconductors with d-wave symmetry. At low doping and low temperatures the effect should be detectable also in loops having a radius hundreds of times larger than the superconducting coherence length, like in Ring A. Nevertheless according to the same theory this *h/e* component should be vanishingly small approaching $T_C$ thus negligible in our system. Moreover, as far as we know, these theories would not explain, in any temperature and doping range, the suppression of *h/2e* signal at the lowest temperature as observed in our data.

## 4. Conclusions

We have shown a transition of an underdoped NBCO nanostructure from a high temperature (T> 2-3 K) state, where Cooper pair transport is allowed, to a low temperature state, where Cooper pair flow is inhibited because of dramatic suppression of proximity effect. The combined analysis of magnetoconductance and conductance measurements points to the activation of a strong pair breaking mechanism at T <2 K, strongly suggestive of the formation of magnetic domains in the underdoped phase of cuprates at lower temperatures. We also demonstrated coherent single charge transport in underdoped normal NBCO at 250 mK and at length scales at least comparable to those of Ring A.

**Acknowledgments**

The Authors would like to thank P. Pingue and V. Piazza for a critical reading of the manuscript and P. Lucignano for useful discussions. DS, PP and FT acknowledge the support ESF projects 'MIDAS' and 'NES'.

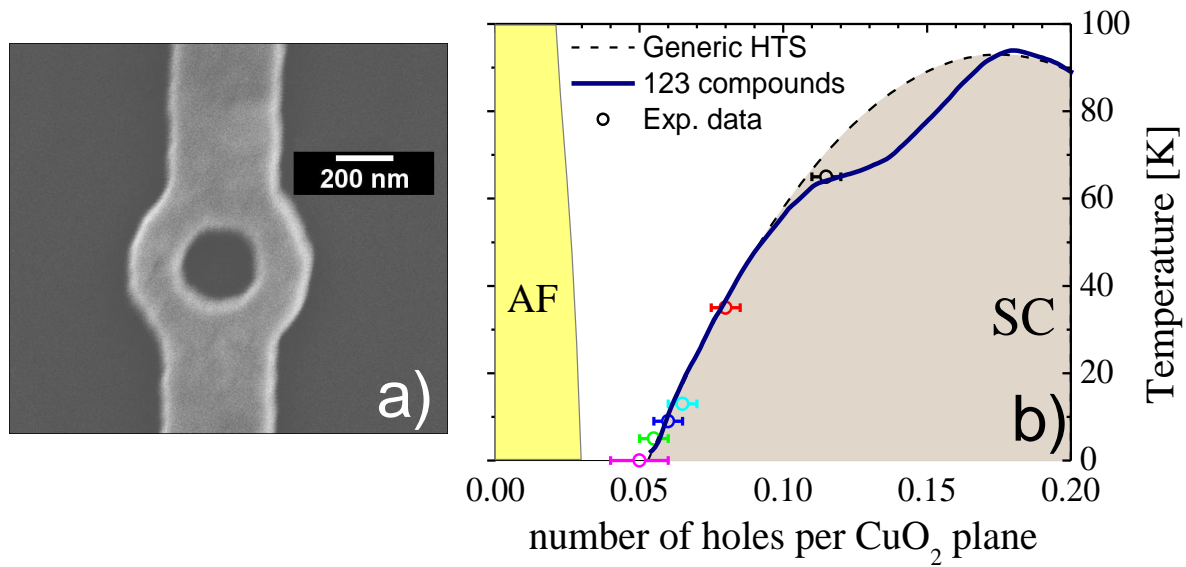

**Figure 1.** a): Scanning electron micrograph of Ring A. b): Typical phase diagram for cuprates (dashed line) and 123 compounds (solid line). Open circles are experimental values of $T_C$ and doping for: as grown film (black), oxygen reduced film (red), leads of Ring B (Cyan), leads of Ring A (dark blue), Ring B (green), Ring A (pink). The doping of ring A was extrapolated from the other experimental data.

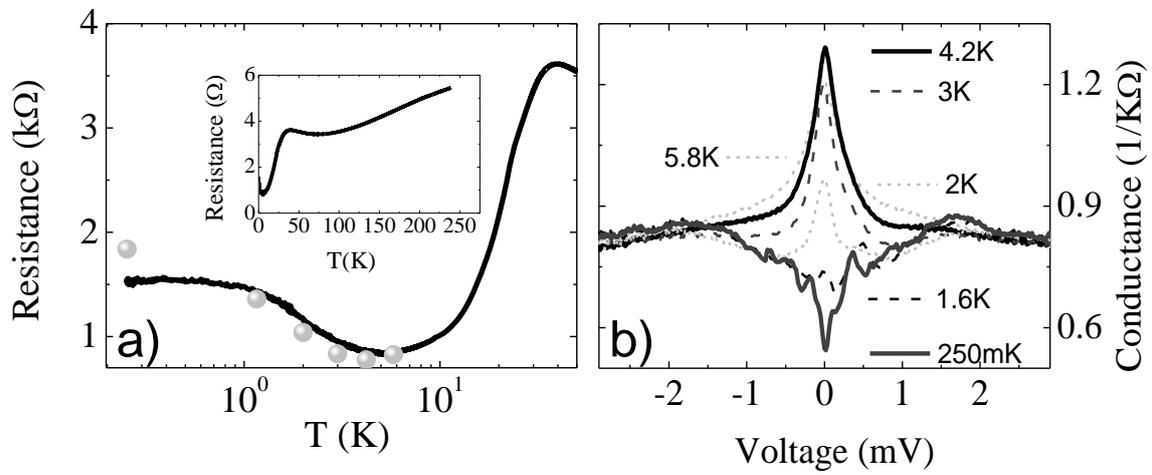

**Figure 2.** a): Resistance vs. Temperature curve (black line) of Ring A taken using 100nA bias current and zero bias resistance (filled circles) taken from dG(V) curves (panel on the right). inset: R vs. T up to high temperatures. b): differential conductance vs. voltage from 5.8K to 250mK of Ring A.

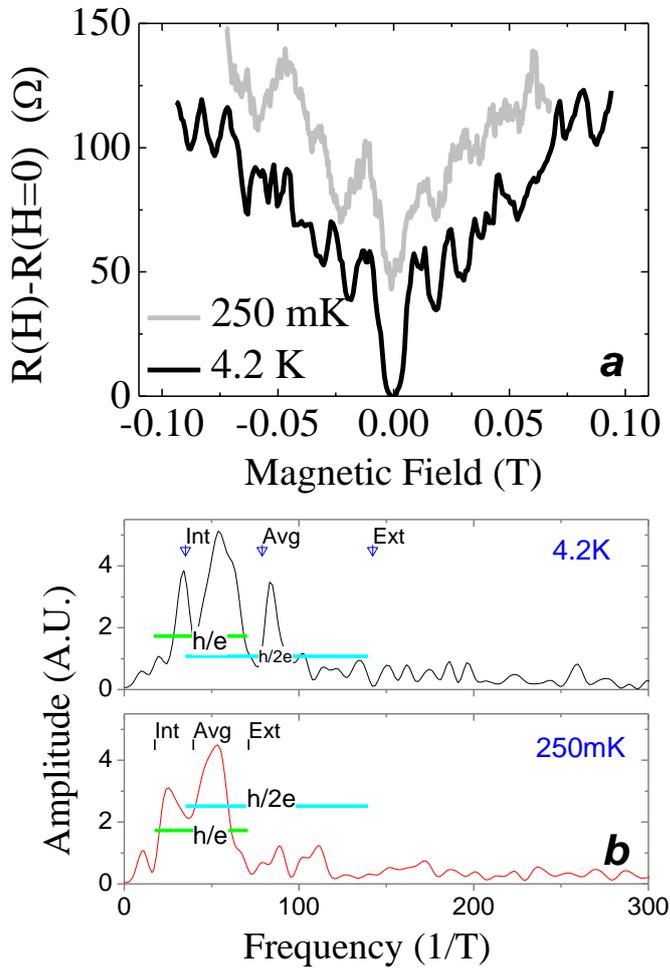

**Figure 3.** a) Zero bias resistance vs magnetic field at 4.2K and 250mK for Ring A. The curve at 250 mK has been translated upward of 50 Ω for clarity. b) Fast Fourier Transform of the two curves in a) after a subtraction of parabolic background due to single wire magnetoresistance. h/e and h/2e bars indicate the frequency (1/T) range calculated from ring sizes and assuming h/e and h/2e flux periodicity respectively (see text). Arrows and lines indicate internal, average and external radius of the ring calculated supposing h/2e or h/e respectively. At both temperatures the main peaks of the FFT are at the lower frequencies of h/2e bar and lie partly outside. In contrast the peaks fill the whole size of h/e bar.

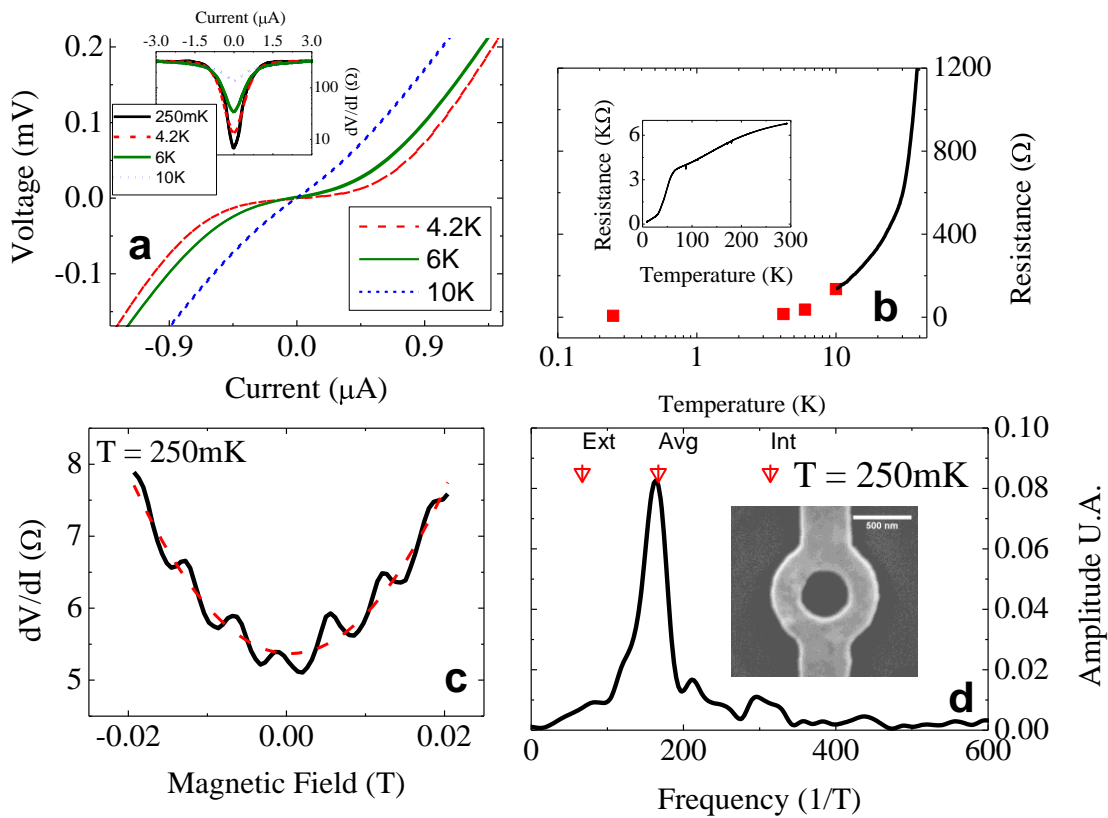

**Figure 4** Transport measurements for Ring B: a) IV curves from 4,2K to 10K. Inset shows dV/dI vs I in semilog scales. The small residual resistance (few ohms) on the supercurrent branch persists down to 250mK and is likely related to thermally activated flux flow in the nanoring and/or in the electrodes. b) Resistance vs. Temperature. c) Zero bias resistance vs. Magnetic Field d) FFT transform of data in panel c) frequencies corresponding to the average, internal and external radius have been calculated considering an h/2e periodicity. Inset: sem image of Ring B, scale bar corresponds to 500nm.